\documentstyle[12pt,epsf,rotate]{article}
\def\be{\begin{equation}}					 
\def\ee{\end{equation}}
\def\ber{\begin{eqnarray}}
\def\eer{\end{eqnarray}}	
\def\dint{\mathop{\intop\kern-0.5em\intop}}

\begin{document}
\vspace*{2cm}
\begin{center}
{\bf ON A MANIFESTATION OF DIBARYON RESONANCES \\IN THE
STRUCTURE OF PROTON-PROTON TOTAL\\ CROSS-SECTION AT LOW
ENERGIES} \\

\vspace{4mm}
{A.A. Arkhipov\\
{\it Institute for High Energy Physics \\
 142280 Protvino, Moscow Region, Russia}}\\
\end{center}
\vspace{4mm}

\centerline{\bf Abstract}
\vspace{4mm}
A manifestation of narrow diproton resonances in the early discovered
global structure of  proton-proton total cross section (see
\cite{8,9}) at low energies is discussed. It is also discussed the
existence of new particle with the mass $1.833\,MeV$
predicted early.

\section{Introduction}
\noindent
We are all know an incessant interest to the physics of dibaryons and
this session at the Conference is an additional confirmation of that
\cite{1,2,3}.  It is well known fact that diprotons have
experimentally been observed as a narrow structures in the
distributions over invariant mass of proton-proton system in the
processes of proton-nucleus interaction \cite{4,5,6}. The physical
origin of these narrow structures is high interest because it has
fundamental importance which is related to the nature of fundamental
nucleon-nucleon forces and not only to this one. However the
experimental and theoretical understanding of the dibaryon physics is
far from desired. 

From experimental point of view it would very well done to obtain a
strong statement concerning the observation of narrow dibaryons
regardless of their origin. However at present time there are many
experiments where we can find quite an opposite results: some authors
state that they have observed such narrow dibaryons \cite{4,5,6} but
the others make the contrary conclusion. It is usually supposed that
the main reason for these disagreements is the weakness of dibaryons
signatures compared to the physical background of a given process
under an experimental study. In that case there are needed the
experiments with a high precision. Of course we can always explain a
contradiction between the different experiments by a poor energy
resolution and statistics or by kinematically unfavourable conditions
and all somethings like that. Therefore it would be very desirable to
have the measurements with one and the same positive signals coming
from different kinds of experiments.

Certainly it's bad that at present time there is no theory which can
explain the existence of the dibaryons and describe them.

I will adress here a nontrivial physical phenomenon related to a
manifestation of diproton resonances in the proton-proton total cross
sections at low energies. We faced with the phenomenon in our
study of global structure for the nucleon-nucleon total cross
section.

Mabe it should be emphasized a common experimental point of view that
the total cross section is not a suitable characteristic to study the
resonance physics. Nevertheless we will show that the existing
experimental data set on proton-proton total cross sections allowed
us to find a clear signatures for diproton resonances. Let me remind
you what was the beginning on.

\section{Global structure of proton-proton(antiproton) total cross
sections}

Recently a simple theoretical formula describing the global structure
of $pp$ and  $p\bar p$ total cross-sections in the whole range of
energies available up today has been derived. The fit to the
experimental data with the formula was made, and it was shown that
there is a very good correspondence of the theoretical formula to the
existing experimental data obtained at the accelerators \cite{7,8}. 

Moreover it turned out there is a very good correspondence of the
theory to all existing cosmic ray experimental data as well. The
predicted values for  $\sigma_{tot}^{pp}$ obtained
from theoretical description of all existing accelerators data are
completely compatible with the values obtained from cosmic ray
experiments \cite{9}. The global structure of proton-proton total
cross section is shown in Fig. 1 extracted from paper \cite{9}.

\begin{figure}[t]
\begin{center}
\begin{picture}(288,188)
\put(15,10){\epsfbox{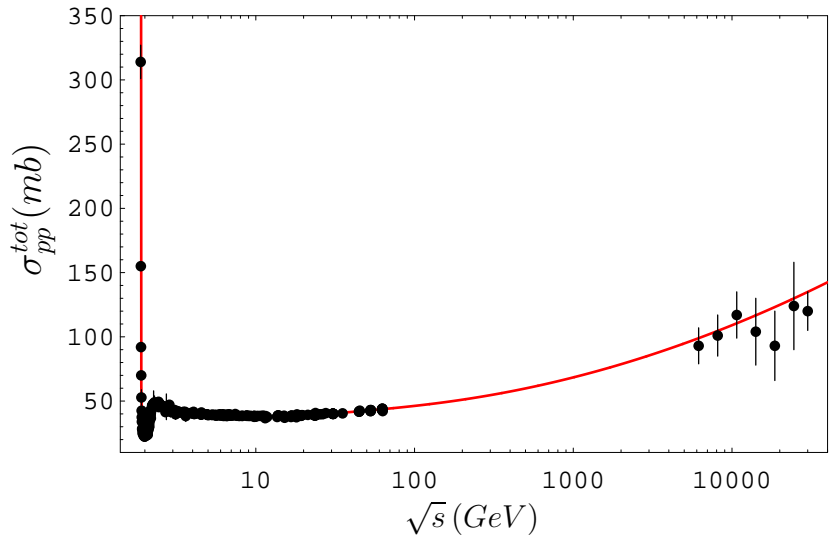}}
\put(144,0){$\sqrt{s}\, (GeV)$}
\put(-5,95){\rotate{\large$\sigma^{tot}_{pp} (mb)$}}
\end{picture}
\end{center}
\caption[]{\protect{The proton-proton total cross-section versus
$\sqrt{s}$ with the cosmic rays data points from Akeno Observatory
and Fly's Eye Collaboration. Solid line corresponds to our theory
predictions.}}
\label{fig:Fig.1}
\end{figure}

The theoretical formula describing the global structure of
proton-proton total cross section is written below
\[
\sigma_{pp}^{tot}(s) = \sigma^{tot}_{asmpt}(s) 
\left[1 + \left(\frac{c_1}{\sqrt{s-4m^2_N}R^3_0(s)} -
\frac{c_2}{\sqrt{s-s_{thr}}R^3_0(s)}\right)\left(1 + d(s)\right) +
Resn(s)\right],
\]
\[
R^2_0(s) = \left[0.40874044 \sigma^{tot}_{asmpt}(s)(mb) -
B(s)\right](GeV^{-2}),
\]
\[
\sigma^{tot}_{asmpt}(s) = 42.0479 + 1.7548 \ln^2(\sqrt{s}/20.74),
\]
\[
B(s) = 11.92 + 0.3036 \ln^2(\sqrt{s}/20.74),
\]
\[
c_1 = (192.85\pm 1.68)GeV^{-2},\quad c_2 = (186.02\pm 1.67)GeV^{-2},
\]
\[
s_{thr} = (3.5283\pm 0.0052)GeV^2,
\]
\[
d(s) = \sum_{k=1}^{8}\frac{d_k}{s^{k/2}},\quad Resn(s) =
\sum_{i=1}^{N}\frac{C_R^i s_R^i
{\Gamma_R^i}^2}{\sqrt{s(s-4m_N^2)}[(s-s_R^i)^2+s_R^i{\Gamma_R^i}^2]}.
\]
For the numerical values of the parameters $d_i (i=1,...8)$ see
original paper \cite{8}. 
\begin{table}[hbt]
\caption{Diproton resonances.}\label{tab}
\begin{center}
\begin{tabular}{|l|c|c|r|}\hline   
$ m_R(MeV) $ & $\Gamma_R(MeV) $ & Reference & $C_R(GeV^2)$  \\ \hline     
$ 1937\pm 2 $ & $ 7\pm 2 $ & \cite{6} & $ 0.058\pm 0.018 $ \\ 
$ 1947(5)\pm 2.5 $ & $ 8\pm 3.9 $ & \cite{4}& $ 0.093\pm 0.028 $\\ 
$ 1955\pm 2 $ & $ 9\pm 4 $ & \cite{6} & $ 0.158 \pm 0.024 $ \\
$ 1965\pm 2 $ & $ 6\pm 2 $ & \cite{6} & $ 0.138 \pm 0.009 $ \\ 
$ 1980\pm 2 $ & $ 9\pm 2 $ & \cite{6} & $ 0.310 \pm 0.051 $ \\
$ 1999\pm 2 $ & $ 9\pm 4 $ & \cite{6} & $ 0.188\pm 0.070 $ \\ 
$ 2008\pm 3 $ & $ 4\pm 2 $ & \cite{6} & $ 0.176 \pm 0.050 $ \\ 
$ 2027\pm ? $ & $ 10 - 12 $ &  & $ 0.121\pm 0.018 $ \\ 
$ 2087\pm 3 $ & $ 12\pm 7 $ & \cite{6} & $ -0.069\pm 0.010 $ \\ 
$ 2106\pm 2 $ & $11\pm 5 $ & \cite{6} & $-0.232 \pm 0.025 $ \\ 
$ 2127(9)\pm 5 $ & $ 4\pm 2 $ & \cite{6} & $ -0.222\pm 0.056 $ \\ 
$ 2180(72)\pm 5 $ & $ 7\pm 3 $ & \cite{6} & $ 0.131\pm 0.015 $ \\ 
$ 2217\pm ? $ & $ 8 - 10 $ &  & $ 0.112\pm 0.031 $ \\ 
$ 2238\pm 3 $ & $22\pm 8 $ & \cite{6} & $ 0.221 \pm 0.078 $ \\ 
$ 2282\pm 4 $ & $24\pm 9 $ & \cite{6} & $ 0.098 \pm 0.024 $ \\
\hline
\end{tabular}
\end{center}
\end{table}
It should be pointed out that the
mathematical structure of the formula is very simple and physically
transparent: the total cross section is represented in a factorized
form. One factor describes high energy asymptotics of total cross
section and it has the universal energy dependence predicted by the
general theorems in local quantum field theory (Froissart theorem).
The other factor is responsible for the behaviour of total cross
section at low energies and it has a complicated resonance
structure.  However this factor has also the universal asymptotics
at elastic threshold. It is a remarkable fact that the low energy
asymptotics of total cross section at elastic threshold is dictated
by high energy asymptotics of three-body (three-nucleon in that case)
forces. The appearance of new threshold $s_{thr}=3.5283\, GeV^2$ in
the proton-proton channel, which is near the elastic threshold, is
nontrivial fact too. 

Some experimental information concerning the diproton resonances is
collected in Table 1.
The positions of resonances and their widths, listed in Table 1, were
fixed in our fit, and only relative contributions of the resonances
$C_R^i$ have been considered as free fit parameters. Fitted
parameters $C_R^i$ obtained by the fit are listed  in Table 1 too. It
should be remarked that the experimental data set on proton-proton
total cross sections revealed the existence of two unknown resonances
with the masses $\sim 2027\,MeV$ and $\sim 2217\,MeV$. These
resonances were also included on our fit. Some known diproton
resonances are not included in the list by the reason of our computer
allowance. We plan to make a more extended analysis in the future.

\begin{figure}[t]
\begin{center}
\begin{picture}(288,188)
\put(15,10){\epsfbox{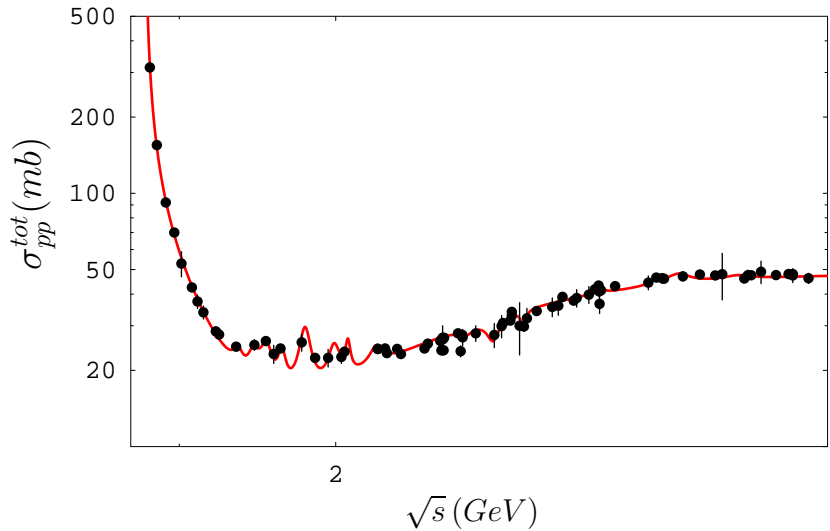}}
\put(144,0){$\sqrt{s}\, (GeV)$}
\put(-5,95){\rotate{\large$\sigma^{tot}_{pp} (mb)$}}
\end{picture}
\end{center}
\caption[]{\protect{The proton-proton total cross-section versus
$\sqrt{s}$ at low energies. Solid line corresponds to our theory
predictions.}}
\label{fig:Fig.2}
\end{figure}

\begin{figure}[t]
\begin{center}
\begin{picture}(288,188)
\put(15,10){\epsfbox{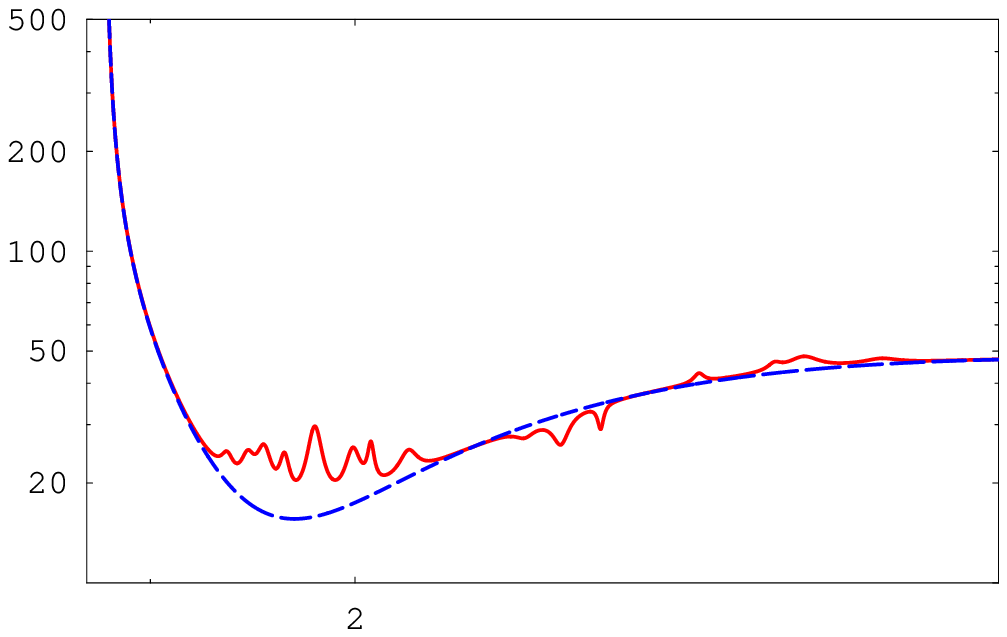}}
\put(144,0){$\sqrt{s}\, (GeV)$}
\put(-5,95){\rotate{\large$\sigma^{tot}_{pp} (mb)$}}
\end{picture}
\end{center}
\caption[]{\protect{The resonance structure for the proton-proton
total cross-section versus $\sqrt{s}$ at low energies. Solid line is
our theory predictions. Dashed line corresponds to the ``background"
where all resonances are switched off.}}
\label{fig:Fig.3}
\end{figure}
Our fitting curve is shown in Fig. 2. 
We also plotted in Fig. 3 the resonance structure for proton-proton
total cross section at low energies without the experimental points
but with dashed line corresponding the ``background" where all
resonances are switched off. As it is seen from this Figure there is
clear signature for the diproton resonances.

\section{Conclusion}

\begin{itemize}
\item {It appears the diproton resonances are confirmed by the
data set for proton-proton total cross section at low energies from
statistical point of view (good fit!).}

\item {There is a big bag (``bol'shoi korob") with many dibaryon
resonances. This korob (bag) is not completely filled yet till now!
How many dibaryon resonances are there?}

\item {There are many questions???...There are no answers!!!...What
is the physical nature and dynamical origin of dibaryon resonances?
What are the quantum numbers: spin, isospin, and so on. A nontrivial
fact in our fitting games is the observation that three resonances
with the mass 2087, 2106, 2127 $MeV$ have an odd parity.}

\item {Without any doubt the physics of dibaryon resonances is very
interesting, very exciting, very promising, very..., very... part of
elementary particle and nuclear physics.}

\item {From the global structure it follows that new threshold, which
is near the elastic one, looks like a manifestation of a new unknown
particle:}

\[
\sqrt{s_{thr}} = 2 m_p + m_{\cal L},\qquad m_{\cal L} = 1.833\pm
0.001\,MeV.
\]
We predicted the position of new threshold with a high accuracy.
\vspace{1mm}
\item {It seems $\cal L$-particle may have many faces. We could take
a refreshing thought that $\cal L$-particle may be a bound state of
photons--``photoball", or a bound state of electron-positron pairs
embedded in continuum, or very deeply bounded system of pions. A very
intriguing idea that $\cal L$-particle is Higgs particle, which is
well known theoretically but it is not observed experimentally, is
admissible one as well. Is $\cal L$-particle a photonium,
positronium, pionium, and so on x-onium?}

\item {Could one make an experiment to search $\cal L$-particle? It
is very probably that $\cal L$-particle has been observed in
Darmstadt. We find in the abstract of paper \cite{10}: ``The most
pronounced line appears at a sum  energy of $\sim 810\,keV$,
corresponding to an invariant mass of $\sim 1.83\,MeV/c^2$." This
result was confirmed by the other group a year later \cite{11}. Now
we can understand an independence of Darmstadt effect on the content
of beam and target nuclei because this is a manifestation of
fundamental nucleon-nucleon dynamics. It's a pity, the present status
of Darmstadt efect is not so stable. That is why, it would be very
desirable to make new experiments to search $\cal L$-particle.}

\item {Could one measure a missing mass spectra in one-particle
$pp\rightarrow pX$ and in two-particle $pp\rightarrow ppX$
inclusive reactions with a high precision and with a high resolution
in missing mass?

Such measurements will shed more light on the questions surrounding
the nature of diproton resonances.}

\item {Surely, it is very important to perform systematic studies
and precise calculations using quantum field theoretical methods. In
this respect we hope that the discovery of quasicrystal structure of
the vacuum in quantum field theory \cite{12} will help us to
understand the new sites of the fundamental dynamics.}

\end{itemize}

\section*{Acknowledgements}

I am grateful to A.M. Zaitsev provided me the possibility to attend
the Conference HADRON 2001 and to present the Report there.

\end{document}